\documentclass[twocolumn,showpacs,preprintnumbers,amsmath,amssymb,floatfix]{revtex4}

\usepackage{graphicx}
\usepackage{dcolumn}
\usepackage{bm}

\newcommand{\pom}{\bf I\! P}

\begin{document}

\title{Diffractive production of dijets by \\
double Pomeron exchange processes}

\author{R. J. M. Covolan}
\author{M. S. Soares}
\altaffiliation{Presently at DESY, Hamburg, Germany.}
\affiliation{Instituto de F\'{\i}sica {\em Gleb Wataghin} \\
{\small Universidade Estadual de Campinas, Unicamp} \\
{\small 13083-970 \ Campinas \ SP \ Brazil }}

\begin{abstract}

A phenomenological description of diffractive dijet hadroproduction via
double Pomeron exchange is presented. This description is based on a modified
version of the Ingelman-Schlein model which includes the evolution of the
Pomeron structure function and corrections regarding rapidity gap suppression
effects. The same quark-dominant Pomeron structure function employed in 
a previous report to describe diffractive dijet and W production via single Pomeron
processes is shown here to yield results consistent with the available data
for double Pomeron processes as well.

\end{abstract}


\maketitle

\section{Introduction}

Since the seminal proposal by Ingelman and Schlein \cite{ingelman}, the study
of hard diffractive processes has become one of the most active and challenging
fields in high energy physics. Diffractive dijet production, in particular,
has been the object of extensive experimental and theoretical analyses via
several processes and in quite different kinematical domains (see, for instance,
Ref. \cite{jung} for a recent summary overview). 

In a recent paper \cite{spe}, we presented an analysis of diffrative 
hadroproduction of dijets and W's via single Pomeron exchange (SPE). 
In the present paper, we apply basically the same picture to analyze dijet 
measurements performed by the Collider Detector at Fermilab (CDF) Collaboration 
\cite{dpe_cdf}, but this time 
via double Pomeron exchange (DPE) processes. Such processes, produced through
quasi-elastic ${\bar p}p$
collisions, are characterized by large rapidity gaps in the fragmentation
regions of both hemispheres and an isolated central hadronic system within
which dijet events are found. A phenomenological view inspired by Regge theory
allows one to conceive these processes as pure Pomeron-Pomeron interactions 
(Fig.~1).

DPE phenomenology has been the focus of intense theoretical investigation
not only for its intrinsic interest in quite different scenarios \cite{varios1},
but also because DPE processes are considered to be one of the main mechanisms
leading to Higgs boson production and new physics signals within a very
clean experimental environment \cite{varios2}. These aspects and the availability 
of experimental data \cite{dpe_cdf} make evident the importance of further phenomenological 
studies on central diffractive dijet production via DPE aiming at describing these 
and more rare events possible to occur by the same reaction channel. 

We show here a description of diffractive dijet production via DPE
by using a modified version of the Ingelman-Schlein (IS) model \cite{ingelman}
in which the Pomeron structure function is evolved according to the
Dokhshitzer-Gribov-Lipatov-Altarelli-Parisi (DGLAP) evolution formalism
\cite{dglap}. The results obtained via IS model are corrected in order to
consider the effect of rapidity gap suppression. This is done
according to the scheme developed by Goulianos in Refs.~\cite{dino,dino2}.


\section{SPE and DPE Cross Sections}

We start by recalling the expressions used to calculate dijet production
via SPE, which are detailed elsewhere \cite{mara}.
From these expressions, the cross section for dijet production via DPE are
readly obtained.

Let us initially consider hadrons $A$ and $B$ colliding and giving rise to
ordinary, {\it nondiffractive} (ND) dijet production. In this case, the cross 
section in terms of the
dijet rapidities ($\eta, \eta'$) and transversal energy $E_T$ is given by:
\begin{eqnarray}
\nonumber
\left(\frac{d\sigma}{d\eta}\right)_{jj}=&&\sum_{\rm partons}
\int_{E_{T\rm{min}}}^{E_{T\rm{max}}}
d E_T^2 \int_{\eta'_{\rm{min}}}^{\eta'_{\rm{max}}} d\eta' \\ 
&&\times\ x_a f_{a/A}(x_a,\mu^2)\ x_b  f_{b/B}(x_b,\mu^2)
\left(\frac{d\hat{\sigma}}{d\hat{t}}\right)_{jj}
\label{modpar}
\end{eqnarray}
where
\begin{equation}
x_a = \frac{E_T}{\sqrt{s}}(e^{-\eta}+e^{-\eta^{\prime}}), \ \ \ \ \ \ \ x_b
= \frac{E_T}{\sqrt{s}}(e^{\eta}+e^{\eta^{\prime}}),
\label{xbj}
\end{equation}
with
\begin{eqnarray}
\ln{\frac{E_T}{\sqrt{s}-E_T\ e^{-\eta }}} \leq \eta ' \leq
\ln{\frac{\sqrt{s}-E_T\ e^{-\eta }}{E_T}}
\end{eqnarray}
and
\begin{eqnarray}
E_{T\ \rm{max}}=\frac{\sqrt{s}}{e^{-\eta }+e^{\eta }},
\label{etmax}
\end{eqnarray}
being that $E_{T\ \rm{min}}$ and the $\eta$ range are determined by
the experimental cuts.

Equations~(\ref{modpar})-(\ref{etmax}) summarize the leading-order QCD 
procedure to
obtain the {\em nondiffractive} dijet cross section (next-to-leading-order
contributions are not essential for the present purposes; see Ref.~\cite{mara}).
In order to obtain the corresponding expression for
SPE processes according to the IS approach, we assume that one of the hadrons, say
hadron $A$, emits a Pomeron whose partons interact with partons of the hadron $B$.
Thus the parton distribution  $x_a f_{A}(x_a, \mu^2)$ in
Eq.~(\ref{modpar}) is replaced by the convolution between a putative
distribution of partons in the Pomeron, $\beta f_{\tt I\! P}(\beta,\mu^2)$,
and the ``emission rate" of Pomerons by $A$, $g_{\pom}(\xi,t)$.
This last quantity, $g_{{\tt I\! P}}(\xi,t)$,
is the so-called Pomeron flux factor whose explicit formulation in
terms of Regge theory is given below. By using this procedure and defining
$g (\xi) \equiv \int_{-\infty}^0 dt\ g_{{\tt I\! P}}(\xi,t)$,
one obtains \cite{mara}
\begin{eqnarray}
\label{convoP}
x_a f_{A}(x_a, \mu^2)\ =\ \int d\xi \
g(\xi)\ {\frac{x_a}{\xi}} f_{\pom}
({\frac{x_a}{\xi}}, \mu^2).
\end{eqnarray}
By inserting the above structure function into Eq.~(\ref{modpar}),
the cross section for diffractive hadroproduction of dijets via single
Pomeron exchange can be expressed as
\begin{widetext}
\begin{eqnarray}
\left(\frac{d\sigma_{SPE}}{d\eta}\right)_{jj}=\sum_{a,b,c,d}
\int_{E_{T_{min}}}^{E_{T_{max}}} dE_T^2
\int_{\eta'_{min}}^{\eta'_{max}} d\eta' 
\int_{\xi_{min}}^{\xi_{max}}
d\xi \ g(\xi)\  \beta_a f_{a/{\tt I\! P}}(\beta_a, \mu^2)
\ x_b f_{b/\bar{p}}(x_b, \mu^2)\
\left(\frac{d\hat{\sigma}_{ab\rightarrow cd}}{d\hat{t}}\right)_{jj},
\label{dsigjato}
\end{eqnarray}
\end{widetext}
where $\beta_a = {x_a}/\xi$, with $x_a$ and $x_b$ given by
Eq.~(\ref{xbj}), and $\xi_{min}$ and $\xi_{max}$
established by experimental cuts.


The process we are interested in, however, is a particular form of diffractive
production in which both $p$ and $\bar{p}$ emit Pomerons
giving rise to dijet production in the central region, accompanied by rapidity
gaps in both hemispheres (Fig.~1).

In terms of the IS model, the reaction that effectively occurs is
$\pom \pom \rightarrow jet\ jet$ and so the respective cross section must
involve two convolution products of flux factor and structure function. From
Eq.~(\ref{dsigjato}), the differential
cross section in terms of variables appropriate for the present analysis becomes
\begin{widetext}
\begin{eqnarray}
\left(\frac{d\sigma_{DPE}}{d\eta_{_{ boost}}}\right)_{jj} = \sum_{a,b,c,d} 
\int 2 d\eta^* \int dE_T^2 \int \beta_a f_{\pom}(\beta_a, \mu^2)\ g_N(\xi_{p})
\ d{\xi_{p}} \int \beta_b f_{\pom}(\beta_b, \mu^2)\ g_N(\xi_{\bar p})\ 
d{\xi_{\bar p}}\ 
\left(\frac{{d{\hat{\sigma}}_{ab\rightarrow cd}}}{d\hat{t}}\right)_{jj},
\label{sig_dpe}
\end{eqnarray}
\end{widetext}
where the new rapidity variables are 
$\eta_{_{\rm boost}} =(\eta + \eta')/2$, $\eta^* = (\eta - \eta')/2$, 
and $\beta_{a(b)} = {x_{a(b)}/\xi_{p(\bar p)}}$, with
$\xi_{p({\bar p})}$ giving the momentum fraction carried by the Pomeron emitted by
the proton (antiproton) vertex. In the flux factors $g_N(\xi_{p,\bar p})$, the 
index $N$ indicates that the normalization procedure described below has
been applied.



An important element of the IS approach is the Pomeron flux factor,
which enters the calculations via Eq.~(\ref{convoP}).
It was originally proposed to be taken from the invariant cross section
of (soft) diffractive dissociation processes as it is given by the triple
Pomeron model \cite{ingelman}. 
Here we apply for $g_{\pom}(\xi, t)$, the
Donnachie-Landshoff parametrization \cite{donna} that, before integration
over $t$, reads
\begin{equation}
g_{\tt I\! P}(\xi,t)=\frac{9{\beta}_{0}^{2}}{4{\pi}^2}
F_{1}^2(t)\ {\xi}^{1-2{\alpha}_{\pom}(t)},
\label{dlf}
\end{equation}
where $F_1(t)$ is the Dirac form factor,
\begin{equation}
F_1(t) = \frac{(4m^2-2.79t)}{(4m^2-t)}\ \frac{1}{(1-\frac{t}{0.71})^2}.
\label{dirac_ff}
\end{equation}
Our choice for the Pomeron trajectory in Eq.~(\ref{dlf}) has been 
$\alpha_{\pom}(t)=1.2 + 0.25\ t$,
which is compatible with both Fermilab Tevatron and DESY ep collider HERA data.

Since the flux factor above violates unitarity, we have employed the
(re)normalization procedure proposed by Goulianos \cite{dino},

\begin{eqnarray}
g_N(\xi) = \frac{g(\xi)}
{\int_{\xi_{min}}^{\xi_{max}}g(\xi)\ d\xi},
\label{erenorm}
\end{eqnarray}
which has been recast in terms
of rapidity gap survival probability in Ref.~\cite{dino2}.



The final element to compose the model, the Pomeron structure function, was 
chosen to be a three-flavor quark
singlet at the initial scale, $\mu_0^2 = 2\ GeV^2$, with the gluon
component being generated by DGLAP evolution. The parametrization
used for the valence quark distribuition bas been
\begin{eqnarray}
\nonumber
\beta \Sigma(\beta,Q^2_0) = && [A_1 \exp(-A_2 \beta^2)+
B_1(1-\beta)^{B_2}]\ \beta^{0.001}\\
&& +\  C_1 \exp[-C_2(1- \beta)^2] (1-\beta)^{0.001},
\end{eqnarray}
which includes different contributions of soft, hard, and superhard profiles
according to
the chosen parameters. The results presented below were obtained with the
following parameters: $A_1$ = 4.75 and $A_2$ = 228.4 for the soft part,
$B_1$ = 1.14 and $B_2$ = 0.55 for the hard one, and finally $C_1$ = 2.87 and
$C_2$ = 100 for the superhard term.
DGLAP evolution of the Pomeron parton densities has been
processed by using the program QCDNUM \cite{qcdnum}.

This parametrization has been successfully employed in a recent study \cite{spe} 
aiming at describing diffractive dijet and W production via SPE (see discussion  
below). Besides the Pomeron structure function described above, some calculations 
for non-diffractive and SPE processes presented below require ordinary parton 
densities. Those were taken from Ref.~\cite{gluck}.


\section{Results and Discussion}

The magnitudes of the different components of the Pomeron structure function 
(soft, hard, and superhard) given above have been established such that not only 
would the integrated cross sections of several processes approximately match  
the experimental data, but also the shape of differential
cross sections would resemble those obtained by the experiments.
The former aspect has already been shown in Ref.~\cite{spe}, the latter is 
presented in this section for both SPE and DPE processes in Figs.~2 and 3. 

Figure 2 shows dijet differential cross sections for SPE processes calculated with
a version of Eq.~(\ref{dsigjato}) in which $d\sigma/d\eta$ is transformed into 
$d\sigma/d\xi$. These results have been obtained with the kinematical constraints 
corresponding to those of the D0 experiment \cite{dzero} in which both central 
($|\eta| < 1.0$) and forward ($|\eta| > 1.6$) jets have been measured for the 
center-of-mass energies 630 and 1800 GeV. The different components of the
Pomeron structure function mentioned above have been established such that the 
$d\sigma/d\xi$ 
cross sections shown here would reproduce the main features presented by the 
corresponding experimental 
$\xi$ distributions (see Fig.~4 of Ref.~\cite{dzero} for comparison). We refer the 
reader to \cite{spe} for results in terms of integrated cross sections.

The normalized dijet rapidity distributions presented in Fig.~3, corresponding to 
ND, SPE, and DPE processes, were constructed assuming that 
antiprotons are coming from the right-hand side (RHS) while protons from the left-hand 
side (LHS). For the ND case, we have a symmetrical distribution peaked at 
the central region with no gaps, as it is supposed to be. For SPE processes, 
antiprotons are quasielastically scattered, keeping around 90\% or more of their 
momenta. Slow partons from the Pomeron (emitted by ${\bar p}$) interact with 
on average much faster partons from the protons, giving rise to a dijet 
distribution shifted to the RHS hemisphere and leaving a rapidity gap on the LHS 
hemisphere, where the antiproton is detected. For DPE processes, we have rapidity 
gaps in both sides, as expected. In this case, the central distribution is slightly 
shifted to the left because we have applied cuts corresponding to the experiment
\cite{dpe_cdf}, $0.035 < \xi_{\bar p} < 0.095$ and $0.01 < \xi_{p} < 0.03$, and
consequently the distribution is boosted towards the antiproton fragmentation 
region.

Although these theoretical distributions are not appropriate for direct comparison 
with the available experimental information, we notice that they bear great 
resemblance with the CDF data (see Fig.~3(b) of Ref.~\cite{dpe_cdf}).

Finally, in Table I, we show that our prediction for the DPE cross section, obtained by
integrating Eq.~(\ref{sig_dpe}) within the experimental limits, is below but close 
to the upper limit established in Ref.~\cite{dpe_cdf}.

\begin{table}
\begin{center}
\begin{tabular}{lcr}
\hline\hline
   & CDF   &  Present Analysis\\
\hline
${\sigma}_{DPE}$ ($E_T > 7$ GeV) &  $~~~~~~~ < 3.7$ nb~~~~~~~~ & 2.3 nb~~~~~~~~\\
\hline\hline
\end{tabular}
\caption{\label{table} \sf Integrated cross section for dijet production via 
double Pomeron
exchange. The model prediction by the present analysis is compared with the upper 
limit for this process established  by the CDF Collaboration \cite{dpe_cdf}. }
\end{center}
\end{table}

In summary, we have presented here a variation of the Ingelman-Schlein model 
which allows one to obtain a reasonable description of the available experimental 
information on diffractive dijet production via double Pomeron processes. 
Taken as a whole, the results presented above and those shown in Ref.~\cite{spe}
lend a considerable support to the picture presented here in spite of the existing 
theoretical objections \cite{collins}.


\section{Acknowledgments}

We would like to thank the Brazilian governmental agencies CNPq and FAPESP
for their financial support.


\section{Figure Captions}

Fig.~1 - Schematic view of hadrons A and B interacting via double Pomeron exchange 
and giving rise to dijets in the central region.

\vspace{1cm}

Fig.~2 - Dijet differential cross sections for single Pomeron processes
corresponding to the constraints of the D0 experiment \cite{dzero}. 
Forward jets correspond to $|\eta| > 1.6$ and central jets to  $|\eta| < 1.0$.
The corresponding center-of-mass energy is specified in each figure.

\vspace{1cm}

Fig.~3 - Theoretical predictions of dijet rapidity distributions for 
non-diffractive (ND), single Pomeron exchange (SPE), and double Pomeron exchange 
(DPE) processes (see text for details).

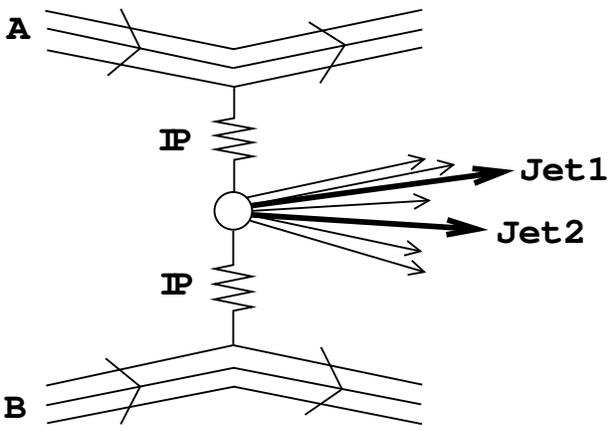
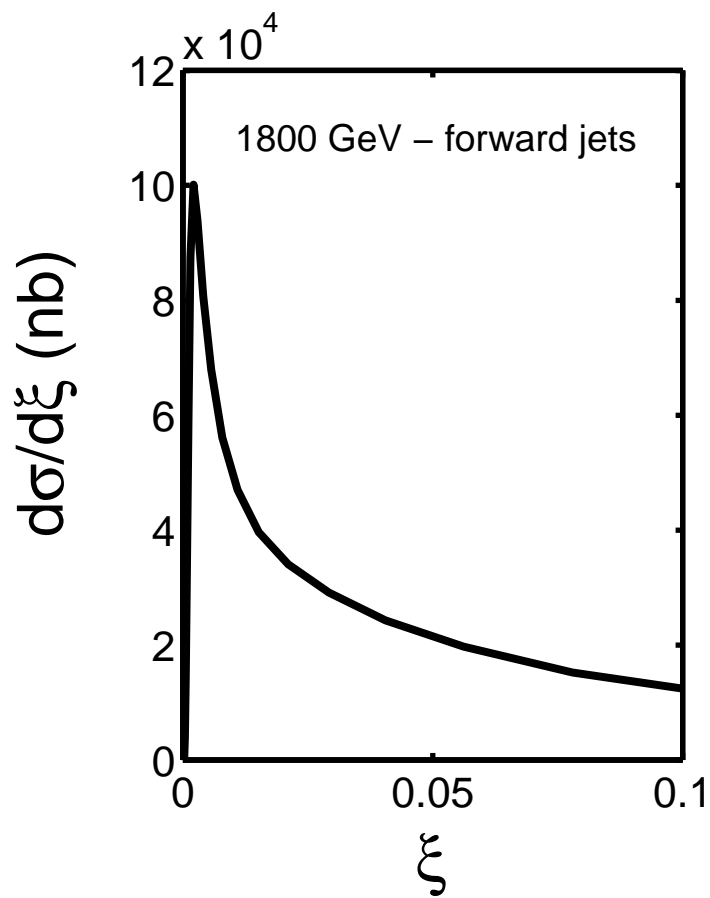
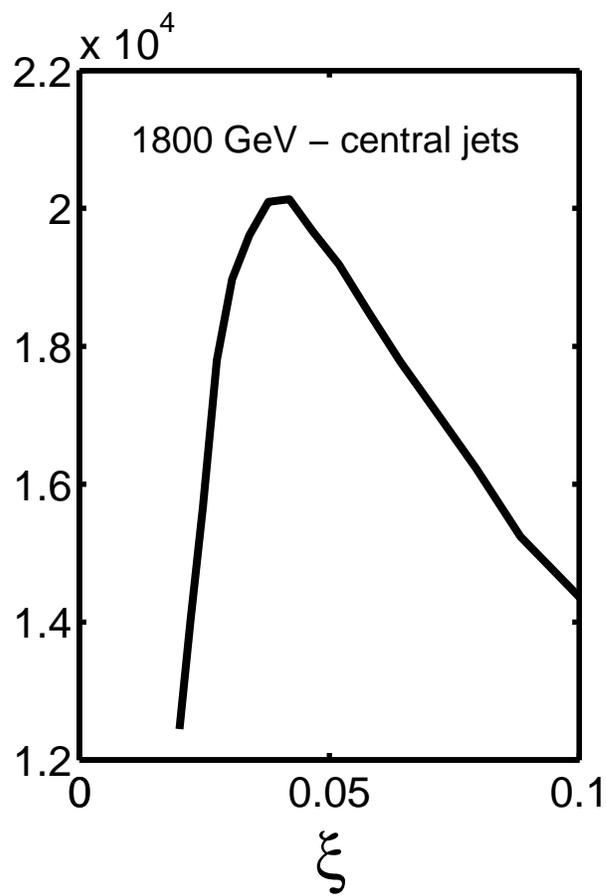
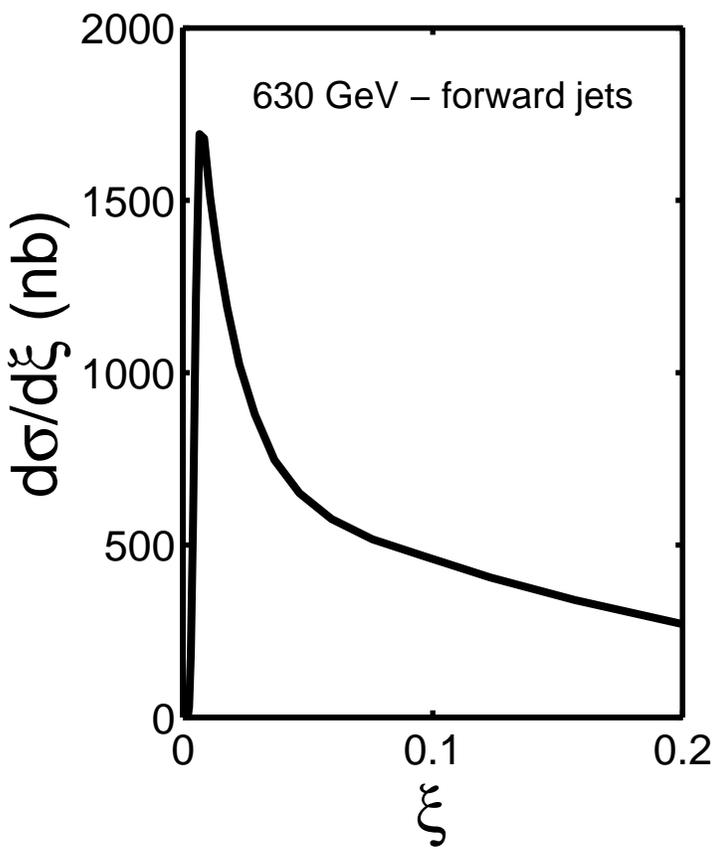
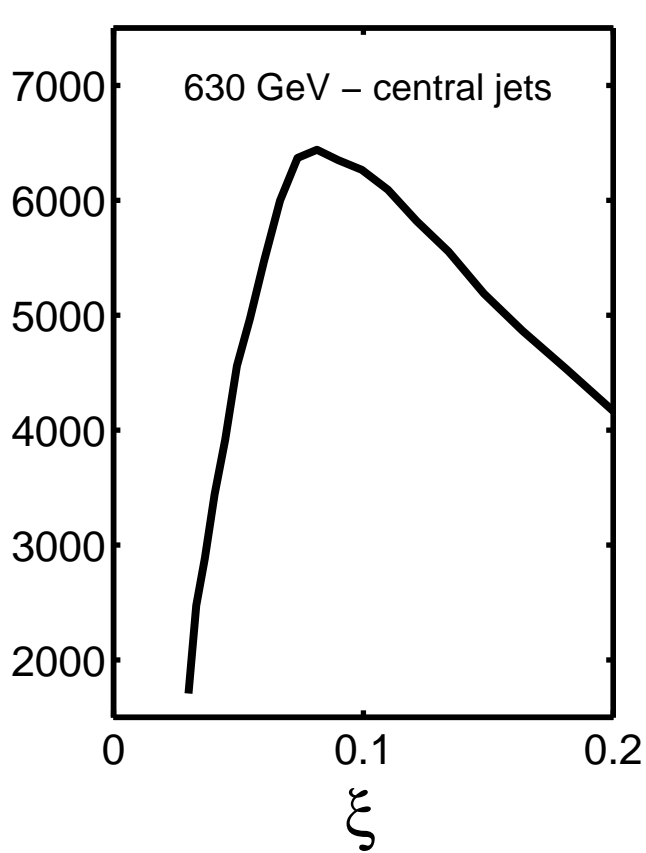
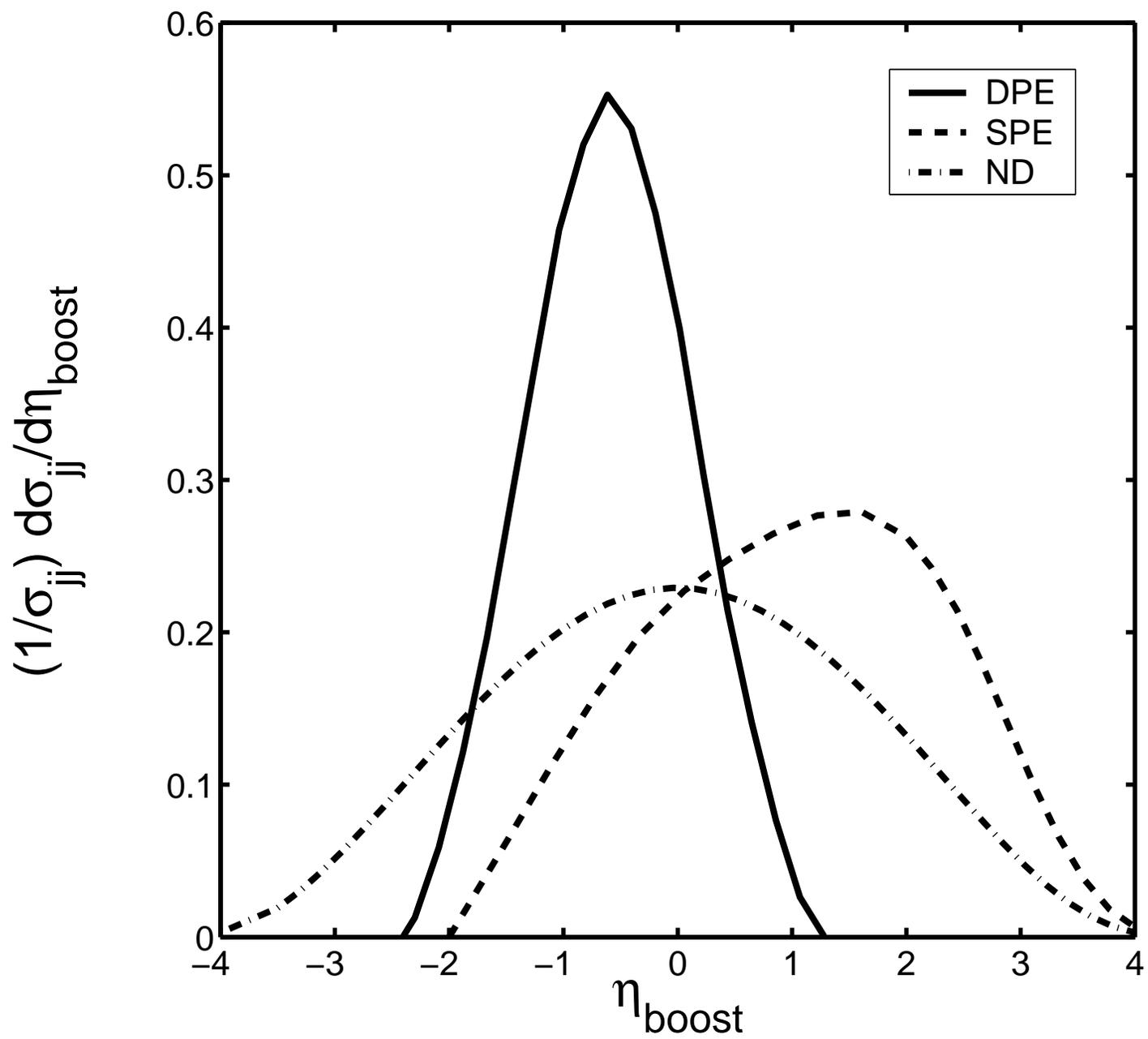
\end{document}